\documentclass[sigconf]{acmart}
\AtBeginDocument{%
  }

\copyrightyear{2025}
\acmYear{2025}
\setcopyright{acmlicensed}
\acmConference[IDC '25]{Interaction Design and Children}{June 23--26, 2025}{Reykjavik, Iceland}
\acmBooktitle{Interaction Design and Children (IDC '25), June 23--26, 2025, Reykjavik, Iceland}
\acmDOI{10.1145/3713043.3728849}
\acmISBN{979-8-4007-1473-3/2025/06}

\begin{document}

\title[Youth as Advisors in Participatory Design]{Youth as Advisors in Participatory Design: Situating Teens’ Expertise in Everyday Algorithm Auditing with Teachers and Researchers}

\author{Daniel J. Noh}
\orcid{0009-0002-7219-1988}
\email{dnoh@upenn.edu}
\affiliation{%
  \institution{University of Pennsylvania}
  \city{Philadelphia}
  \state{Pennsylvania}
  \country{USA}
}

\author{Deborah A. Fields}
\orcid{0000-0003-1627-9512}
\email{deborah.fields@usu.edu}
\affiliation{%
  \institution{Utah State University}
  \city{Logan}
  \state{Utah}
  \country{USA}
}

\author{Luis Morales-Navarro}
\orcid{0000-0002-8777-2374}
\email{luismn@upenn.edu}
\affiliation{%
  \institution{University of Pennsylvania}
  \city{Philadelphia}
  \state{Pennsylvania}
  \country{USA}
}

\author{Alexis M. Cabrera-Sutch}
\orcid{0009-0005-6510-0824}
\email{asutch@fi.edu}
\affiliation{%
  \institution{The Franklin Institute}
  \city{Philadelphia}
  \state{Pennsylvania}
  \country{USA}
}

\author{Yasmin B. Kafai}
\orcid{0000-0003-4018-0491}
\email{kafai@upenn.edu}
\affiliation{%
  \institution{University of Pennsylvania}
  \city{Philadelphia}
  \state{Pennsylvania}
  \country{USA}
}

\author{Dana{\'e} Metaxa}
\orcid{0000-0001-9359-6090}
\email{metaxa@seas.upenn.edu}
\affiliation{%
  \institution{University of Pennsylvania}
  \city{Philadelphia}
  \state{Pennsylvania}
  \country{USA}
}

\renewcommand{\shortauthors}{Noh et al.}

\begin{abstract}
    Research on children and youth’s participation in different roles in the design of technologies is one of the core contributions in child-computer interaction studies. Building on this work, we situate youth as advisors to a group of high school computer science teacher- and researcher-designers creating learning activities in the context of emerging technologies. Specifically, we explore algorithm auditing as a potential entry point for youth and adults to critically evaluate generative AI algorithmic systems, with the goal of designing classroom lessons. Through a two-hour session where three teenagers (16--18 years) served as advisors, we (1) examine the types of expertise the teens shared and (2) identify back stage design elements that fostered their agency and voice in this advisory role. Our discussion considers opportunities and challenges in situating youth as advisors, providing recommendations for actions that researchers, facilitators, and teachers can take to make this unusual arrangement feasible and productive.
\end{abstract}



\maketitle

\section{Introduction}
As the field of child-computer interaction (hereafter: CCI) is evolving, there is a growing need to expand participatory design (hereafter: PD) roles for youth---beyond simply using, testing, or designing interactive technologies alongside adults. While different roles have been proposed for including young people as users, testers, informants, and design partners ~\cite{druin2002role}, the role of expert advisor has been mostly reserved for adults with prior content knowledge and teaching experience ~\cite{lehnert2022cci}. Youth's extensive experiences with emerging technologies, however, challenge this presumed hierarchy of expertise ~\cite{vanmechelen2022emerging}. This is especially clear in the context of social media platforms such as TikTok and YouTube---heavily frequented by youth and much less so by adults ~\cite{anderson2023teens}. Here, youth have rich experience interacting with algorithmic systems such as generative AI filters and recommendation systems, observing outputs, and sensemaking about faults and biases in those systems ~\cite{anderson2023teens}. As researchers and teachers develop activities and tools to promote computational empowerment ~\cite{dindler2023dorit}---which involves both creative construction and critical examinations of emerging technologies---we argue for involving youth in more significant roles. Rather than engaging them solely as informants and design partners, we advocate for roles that draw on youth expertise and experience to make these activities and tools more accessible to learners.

But situating youth in meaningful roles (here, situating refers to how we, researcher-designers create a context for legitimate participation) in such PD contexts presents several challenges. For instance, centering young people in the participatory design of classroom learning activities remains difficult, as the experiences of teachers and researchers are often prioritized in the design of curriculum. This requires, as ~\citet{smith2023research} argue, “long-term collaboration with teachers and students to co-develop new learning practices from the bottom up.” A further critical challenge is structuring PD sessions in ways that youth are empowered with the authority and agency to share their ideas, ensuring their voices and opinions are meaningfully integrated into classroom learning activities ~\cite{bodker2017tying, iversen2012scandinavian}. For that reason, we must provide insights into “the back stage of design” ~\cite{bodker2017tying}, that is, the complicated and context-specific aspects---including prior workshops, preparatory meetings, planning of activities, and the complex network of interpersonal relationships---that make PD possible before, between, and after these sessions.

In this paper, we illustrate how youth can be engaged in the role of advisors to teachers and researchers and offer expert guidance about the usability and applicability of teacher-designed learning activities. In doing so, we highlight the back stage design elements, including our long-standing relationships with youth, teachers, and a science center educator that enabled this work ~\cite{bodker2017tying, smith2023research}. In our session, we situated three high school teens (16--18 years) as advisors to a group of high school computer science teacher-designers and researcher-designers who were working on designing learning activities to introduce algorithm auditing activities into their classrooms. Algorithm auditing is a method that offers a potential entry point for youth and adults to evaluate algorithmic systems from the outside-in. This approach involves “repeatedly querying an algorithm and observing its output in order to draw conclusions about the algorithm’s opaque inner workings and possible external impact” ~\cite{metaxa2021auditing}. Everyday end-users can initiate audits with or without expert involvement, identifying harmful bias in algorithmic systems ~\cite{shen2021everyday}. In addition to the teens’ experiences with algorithmic systems in their everyday lives, they had prior experience conducting their own everyday algorithm audits. We leveraged both sources of expertise by situating the teens as advisors during a two-hour advisory session in our PD process. Our analysis addresses the following research questions: (1) What kind of knowledge and practices did youth share when situated as expert advisors? and (2) What deliberate back stage design choices for the session and spontaneous facilitation decisions encouraged the youth to take on the role of advisors? In the discussion section, we review what we learned from involving teens in PD, not as users or testers (though these youth previously were), but as advisors.

This paper proposes the role of advisors for youth in PD, discusses opportunities and challenges in situating youth in this role, and provides recommendations researchers, facilitators, and teachers can take to make this unusual arrangement more feasible and productive through an analysis of our facilitation and back stage design.

\section{Background}
In the following sections, we review the complex engagement with various communities needed to design sustainable CCI interventions, developing learning activities on emerging technology for classrooms. We also introduce our approach to situating youth as advisors in PD and address the challenges of setting up youth as advisors in the design of such learning activities.

\subsection{Designing Sustainable CCI Interventions}
Designing sustainable CCI interventions requires complex long-term engagement with various communities ~\cite{iversen2014sustaining}. This is particularly relevant to the design of learning environments and activities about emerging technologies, where the goal is wide implementation in school classrooms ~\cite{smith2023research, vanmechelen2022emerging}. Here, it is crucial to create PD processes that are community-centered, sustainable, replicable, scalable, and evolving ~\cite{iversen2014sustaining, dantec2013infrastructuring, disalvo2012communities, poderi2018participatory}, where relevant parties\footnote{We use the term “relevant parties” rather than “stakeholders”; see ~\cite{reed2024reimagining}.} are empowered to genuinely participate in design and decision-making processes ~\cite{smith2018participatory}. ~\citet{smith2018participatory} propose that this process involves attending to three dimensions of engagement in PD: scoping, developing, and scaling. In \textit{scoping}, the goal is to create a design space where participants with diverse roles in a project (e.g., teachers, students, researchers, science educators) can be empowered to explore the possibilities of what a learning environment could look like. \textit{Developing} is the socio-technical process of producing interventions. Finally, \textit{scaling} focuses on creating opportunities distributed across relevant parties to maintain, scale, and even adapt interventions to their own contexts. They describe, for example, pilot workshops as \textit{scoping}, field studies and design experiments as \textit{developing}, and teacher training and the development of national curriculum as \textit{scaling}.

In this paper, we focus on the early \textit{developing} dimension of a larger project aimed at the design and implementation of scalable learning activities on algorithm auditing, bringing together youth, teachers, and researchers. For this early development process, we conducted a PD workshop with teachers where they learned from algorithm auditing researchers, conducted their own audits, received advice from youth, observed a PD workshop for youth, and brainstormed possible directions for lesson design. This workshop was designed to bring participants together and imagine what learning about algorithm auditing could look like. In this paper, we particularly focus on a session within the workshop in which youth were situated as advisors.

\subsection{Expanding Roles in Participatory Design}
Over the last three decades, extensive efforts have focused on developing various PD approaches and methods to involve young people in the design process of technologies ~\cite{druin2002role, fails2013methods}. ~\citet{druin2002role} outlined different roles that children can play along a continuum ranging from users to testers to informants to design partners, and at times even move between roles ~\cite{landoni2018comparative}. Building on Druin’s work, ~\citet{iversen2017child} analyzed each of these roles in terms of their objectives, processes, and outcomes, and proposed the role of protagonists, in which children are the main agents of the design process. Young people’s involvement in the PD process varies from users who examine and provide feedback for a product after the design has been completed to testers who are involved throughout the design process. As informants, young people can provide insights at various points in the design process, while as design partners, they are involved in creating solutions as early as the ideation stage. As co-researchers, young people join in the analysis of data from their own interaction with computing systems. Lastly, young people can also act as designers of applications ~\cite{harel1991children, kafai1995minds} and be the protagonists of the design process ~\cite{iversen2017child}.

Teens as advisors expands on these roles---mostly focused on children and not teens---by situating them as experts, a role often occupied by adults, including software designers, curriculum developers, content experts, and teachers, who are brought into a project to advise researchers on their development efforts based on their technical, content, or practical expertise. While both informants and advisors can be asked to provide feedback on design ideas, prototypes, or more finished products and address purposes, including appropriateness, impact, and better usability for a particular age group, "advisors" give stronger priority to expertise. Advisors do not just provide information or perspective (though feedback may include this); they are attributed to having some authority and expertise in a domain that colors the feedback they provide. Advisors may be drawn from youth who previously filled other roles. In situating teens as advisors, we intentionally attribute expertise to teenagers---expertise which adults might not have, as many are not expert users of youth-centered applications or familiar with youth interests ~\cite{costanza-chock2020design}. 

In our study, it was crucial to involve teens as advisors in the design of algorithm auditing learning activities, as they bring unique perspectives and ideas that can both challenge and enhance the work of teachers and researchers. Whereas most algorithm auditing research has been conducted by expert researchers ~\cite{bandy2021problematic}, research on everyday algorithm auditing recognizes that end-users who are non-expert auditors are able to detect harmful algorithmic behaviors and make contributions that would otherwise be unnoticed ~\cite{shen2021everyday, devrio2022toward}. Particularly, research shows that \textit{young people} are capable of identifying harmful behaviors in the technologies they are familiar with ~\cite{solyst2023potential, morales2024youth}. A more recent study with teens found that they were able to conduct an algorithm audit, which included articulating hypotheses, conducting tests, and producing reports, uncovering behaviors that had not been previously studied by expert researchers ~\cite{morales2025learning}. Teens’ everyday experiences with emerging technologies---and, for this study, in conducting informal audits---position them to effectively advise designers of learning activities.

\subsection{Challenges of Situating Youth as Advisors}
As mentioned, assigning teenagers the role of advisors also comes with challenges ~\cite{fails2013methods, iversen2014sustaining, lehnert2022cci}. Most CCI research has emphasized younger children, while working with teenagers in PD remains fairly uncommon, though it has seen increasing attention in recent years ~\cite{cesario2022designing, kaspersen2022high, morales2022reimagining}. This trend is partly attributed to the unique developmental stage and cultural contexts of teens ~\cite{druin2002role, fails2013methods}, including intensive time commitments in school, extracurricular activities, and jobs that may preclude the type of design work often done with children in PD ~\cite{read2011understanding}. Those who have involved teens in PD have found ways to manage the lack of time, for instance, by creating “accelerated” and focused design sessions on specific topics ~\cite{pazmino2015designing} or with time-limited design challenges ~\cite{rose2019participatory}, providing incentives and endorsements to motivate teen participation in design ~\cite{iversen2013understanding, bell2016learning}, developing agentic roles such as research-apprentice that center youth authority and responsibility \cite{chatlani2023teen}, and building on existing relationships with participants \cite{morales2022reimagining} including teens already leading innovative efforts in places like libraries ~\cite{roldan2022disrupting}. 

While a common challenge of PD involves dealing with power differentials among relevant parties to ensure that participant voices are respected and heard \cite{bannon2012design}, this challenge becomes even greater when situating youth in the roles of advisors due to the inherent power differential between teen-advisors and adult researchers. Concerns have been raised about teens’ feeling less capable of providing expert insights in intergenerational design environments ~\cite{poole2013interaction}. This becomes a greater challenge because PD in CCI has historically prioritized short-term isolated design sessions with youth that constrain the development of long-lasting relationships ~\cite{read2016ideas}, which may be required for youth to be empowered to advise on design processes. A critical issue in such efforts is structuring activities and sessions in which youth have authority and agency to share their ideas. For this reason, efforts have focused on relationship building ~\cite{bonsignore2016traversing}. More recent work has expanded the inclusion of teens in the design of learning activities that are implemented in the classroom ~\cite{bodker2017tying}. For instance, ~\citet{coenraad2022using} included teenagers as a relevant party, alongside teachers, administrators, and parents, in the design of culturally relevant computer science curriculum. They found that students contributed project prompts and aesthetics related to their cultural heritage. In setting up the different groups, researchers intentionally placed students away from teachers and parents in some activities to avoid the silencing of student voices---an arrangement which we will challenge and further investigate in our study.

Addressing these challenges requires tying into existing networks and relationships between youth, teachers, and researchers ~\cite{bodker2017tying}. This work of relationship building is often disregarded in CCI research, yet it is key to building sustainable PD initiatives that benefit all relevant parties ~\cite{iversen2013understanding}. This is particularly important in making sure that, before becoming advisors, youth are comfortable participating in PD and knowledgeable about the design process and activities. Addressing these challenges also requires us to think carefully about what happens in “the back stage of design,” addressing both horizontal and vertical aspects of participatory processes ~\cite{bodker2017tying}. Horizontal aspects involve the methods and micro-dynamics of participatory work during PD workshops, while vertical aspects involve how a participatory process requires the integration of relevant parties at different levels of authority (e.g., students, teachers, school administrators, policymakers) ~\cite{bodker2017tying}. In setting the stage for teens to be advisors to teachers and researchers, we must also account for the power differentials that exist, as participants of PD often perceive themselves to be at varying vertical levels of authority ~\cite{bodker2017tying}. In this paper, we address horizontal aspects of the micro-dynamics when youth are situated as advisors of teachers and researchers in designing auditing learning activities while acknowledging the vertical power differential among relevant parties and the back stage work that makes this kind of PD possible.

\section{Methods}
We present a descriptive case study of a PD workshop in which high school teens advised teachers and researchers regarding algorithm auditing learning activities as part of a larger project. The overarching goal of this project is to create a set of lessons for high school computer science courses that engage youth in algorithm auditing as a means of critical and approachable engagement with algorithmic bias. Following the PD model developed by ~\citet{smith2018participatory}, we foresee the project taking place over scoping, development, and scaling phases. In the following sections, we first explain the broader context of the design trajectory of the algorithm auditing activities, then introduce the different participants and their histories with the researcher team, and finally share the design of the two-hour session where the three youth served as advisors to teachers and researchers. 

\subsection{Participants}
The participants of the Teen Advisory Session consisted of people with different capacities and roles: three teens, six public school teachers, a science center educator, and three researchers (Table ~\ref{tab:demographic}). All names used in this paper are pseudonyms.

\subsubsection{Teen Advisors}
All teens who participated in an earlier PD workshop held in November 2023 (described below) were invited by Bri, the science center educator, to join the Teen Advisory Session. Three were available in July 2024 to participate. These teens were enrolled in a four-year-long STEM Stars enrichment program at a science center in a Mid-Atlantic city in the United States, which brought together high school students for weekly STEM workshops and summer camps. They also reported that they were on their phones for an average of 7.5 hours a day, with Mel and Nadine reporting themselves as frequent users of TikTok and Instagram and Eric as a frequent user of Snapchat and Instagram\footnote{Of note, Instagram and Snapchat have similar filters to those on TikTok.}. They noted that they used these apps to make funny and GRWM videos\footnote{GRWM (Get Ready With Me) videos are a popular social media trend in which users record themselves getting ready for the day.}, take and send pictures to friends, or watch fun content. The teens were compensated for their participation in the earlier workshop and the Teen Advisory Session through hourly pay.

\subsubsection{Teacher-Designers}
We recruited the six U.S. public high school teachers from a cohort of teachers who had previously participated in professional development and implementation of an equity-driven, physical computing curricular unit for introductory computer science (CS) classes ~\cite{fields2023supporting, fields2023good}. Five teachers worked in schools with a majority of Black and brown or rural youth, one (Phoebe) in a majority White suburban school, providing a rich diversity of experiences with students, school districts, and regions. All teachers received compensation for their time and travel. The term "teacher-designer" foregrounds the teachers' primary role in the PD session as designers; however, for simplicity and ease of reading, throughout the rest of the paper, we simply use "teachers".

\subsubsection{Researcher-Designers}
The researcher-designers present at the session included one researcher from the West Coast, two researchers from the same Mid-Atlantic city as the science center, and the manager for the STEM Stars program. The manager met the teens in 2022 and had worked with them since, on a weekly basis. They also played a key role in coordinating the meetings between the researchers and the teens. All three researchers had a long-standing collaboration with the teachers to varying extents (Table ~\ref{tab:demographic}). Similarly, Zia and Matías had worked with the STEM Stars program for 10 and 5 years, respectively. As with teacher-designers, throughout the remainder of the paper, we use "researchers" for simplicity.

\begin{table*}[t]
  \caption{Overview of demographics of youth, teacher, and researcher participants}
  \label{tab:demographic}
\begin{tabular}{@{}lllll@{}}
\toprule
\textbf{Teens} &
  \textbf{Ethnicity} &
  \textbf{Gender} &
  \textbf{Grade} &
  \textbf{Prior Experience with Researchers} \\ \midrule
Mel &
  Black &
  Female &
  \begin{tabular}[c]{@{}l@{}}US East Coast (Public)\\ 12th\end{tabular} &
  \begin{tabular}[c]{@{}l@{}}2021 design workshop ~\cite{shaw2023reimagining},\\ 2021 PD workshop ~\cite{morales2022reimagining},\\ 2023 PD workshop ~\cite{vogelstein2025rapid}\end{tabular} \\ \midrule
Eric &
  Black &
  Male &
  \begin{tabular}[c]{@{}l@{}}US East Coast (Public)\\ 11th\end{tabular} &
  \begin{tabular}[c]{@{}l@{}}2022 design workshop ~\cite{morales2024not},\\ 2023 PD workshop ~\cite{vogelstein2025rapid},\\ STEM Stars Ambassador\end{tabular} \\ \midrule
Nadine &
  Black &
  Female &
  \begin{tabular}[c]{@{}l@{}}US East Coast (Private)\\ 12th\end{tabular} &
  \begin{tabular}[c]{@{}l@{}}2021 design workshop ~\cite{shaw2023reimagining},\\ 2023 PD workshop ~\cite{vogelstein2025rapid}\end{tabular} \\ \midrule
\multicolumn{3}{l}{\textbf{Teachers}} &
  \textbf{Teaching Experience} &
  \textbf{Prior Experience with Researchers} \\ \midrule
Angela &
  Asian &
  Female &
  \begin{tabular}[c]{@{}l@{}}US West Coast (Metropolitan)\\ 17 years teaching (9 CS)\end{tabular} &
  \begin{tabular}[c]{@{}l@{}}10 years implementing physical computing curriculum\\ and professional development facilitator\end{tabular} \\ \midrule
Jesse &
  Multiracial &
  Male &
  \begin{tabular}[c]{@{}l@{}}US West Coast (Metropolitan)\\ 22 years teaching (12 CS)\end{tabular} &
  \begin{tabular}[c]{@{}l@{}}8 years implementing physical computing curriculum\\ and professional development facilitator\end{tabular} \\ \midrule
Liam &
  \begin{tabular}[c]{@{}l@{}}Japanese/\\ Mexican\end{tabular} &
  Male &
  \begin{tabular}[c]{@{}l@{}}US West Coast (Metropolitan)\\ 15 years teaching (12 CS)\end{tabular} &
  \begin{tabular}[c]{@{}l@{}}10 years co-designing physical computing curriculum,\\ implementation, and professional development facilitator\end{tabular} \\ \midrule
Phoebe &
  \begin{tabular}[c]{@{}l@{}}Black/\\ Cherokee\end{tabular} &
  Female &
  \begin{tabular}[c]{@{}l@{}}US Mid-Atlantic (Suburban)\\ 6 years teaching (6 CS)\end{tabular} &
  \begin{tabular}[c]{@{}l@{}}4 years implementing physical computing\\ curriculum, participating in professional development\end{tabular} \\ \midrule
Leta &
  \begin{tabular}[c]{@{}l@{}}White/\\ Greek\end{tabular} &
  Male &
  \begin{tabular}[c]{@{}l@{}}US Mid-Atlantic (Rural)\\ 14 years teaching (4 CS)\end{tabular} &
  \begin{tabular}[c]{@{}l@{}}4 years implementing physical computing curriculum,\\ participating in professional development\end{tabular} \\ \midrule
Tessa &
  White &
  Female &
  \begin{tabular}[c]{@{}l@{}}US Mid-Atlantic (Metropolitan)\\ 11 years teaching (4 CS)\end{tabular} &
  \begin{tabular}[c]{@{}l@{}}10 years piloting physical computing and algorithm\\ auditing lessons\end{tabular} \\ \midrule
\multicolumn{3}{l}{\textbf{Researchers}} &
  \textbf{Prior Experience with Youth} &
  \textbf{Prior Experience with Teachers} \\ \midrule
Bri &
  Latinx &
  \begin{tabular}[c]{@{}l@{}}Non-\\ Binary\end{tabular} &
  \begin{tabular}[c]{@{}l@{}}Science center educator,\\ Manager of STEM Stars program\end{tabular} &
  None \\ \midrule
Matías &
  Latinx &
  Male &
  \begin{tabular}[c]{@{}l@{}}Led all teen workshops,\\ 5+ years collaborating with the\\ STEM Stars program\end{tabular} &
  \begin{tabular}[c]{@{}l@{}}4 years collecting data and supporting professional\\ development with teachers\end{tabular} \\ \midrule
Clara &
  White &
  Female &
  None &
  \begin{tabular}[c]{@{}l@{}}10 years co-designing physical computing curriculum,\\ leading and supporting professional development, \\ data collection with all teachers\end{tabular} \\ \midrule
Zia &
  \begin{tabular}[c]{@{}l@{}}Middle Eastern/\\ White\end{tabular} &
  Female &
  \begin{tabular}[c]{@{}l@{}}Supported teen workshops,\\ 10 years collaborating with\\ STEM Stars program\end{tabular} &
  \begin{tabular}[c]{@{}l@{}}10 years supporting professional development of\\ physical computing curriculum and algorithm auditing\end{tabular} \\ \bottomrule
\end{tabular}
\end{table*}

\subsection{Participatory Design Workshops with Teens and Teachers}
\subsubsection{Emergent Auditing Practices (EAP) Workshop with Teens}
Our scoping activities began in November 2023, during an Emergent Auditing Practices (EAP) workshop with high school youth in an informal setting to identify various everyday algorithm auditing practices. Over the course of two Saturdays (8 hours total), we met in person with a group of seven high school youth (10th--12th grade youth) who were part of a four-year-long STEM Stars program. All participants had already engaged in previous PD research studies with our team. 

\begin{figure}[h]
\centering
\includegraphics[width=\linewidth]{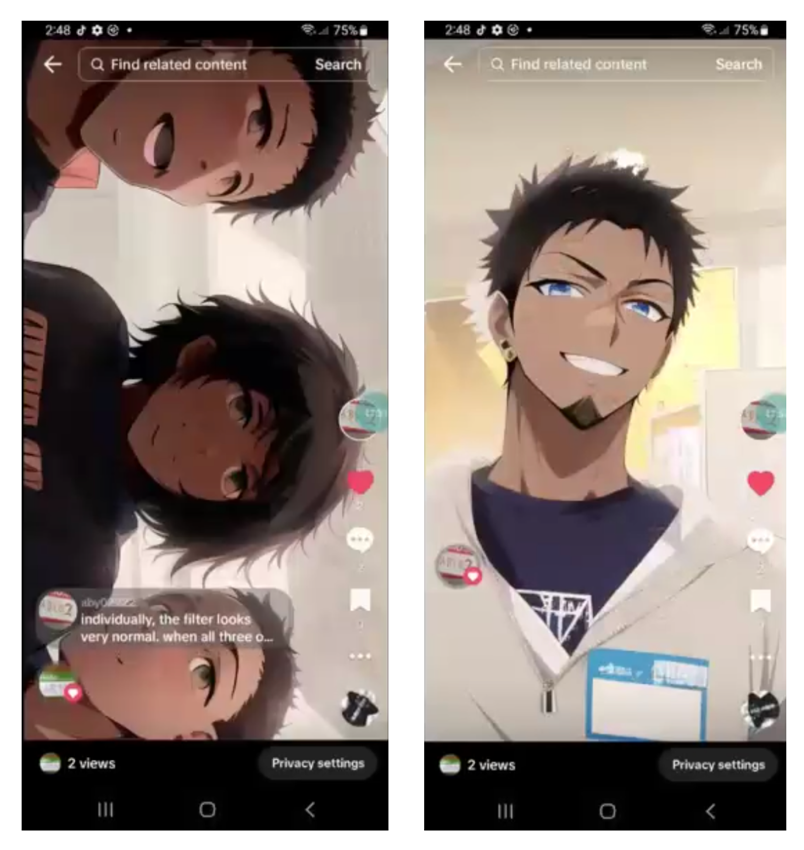}
\Description[Two screenshots of faces altered by a Manga TikTok filter]{Two screenshots of TikTok. Left: two masculine-presenting and one gender-ambiguous face resembling the style of manga. Right: one masculine-presenting face resembling the style of manga.}
\caption{Examples of faces altered by TikTok filters from the EAP workshop.}
\label{fig:tiktok}
\end{figure}

During the Emergent Auditing Practices (EAP) sessions, the youth participants (1) explored their everyday use of AI/ML-powered applications, (2) conducted exploratory auditing activities to evaluate generative AI TikTok filters that transformed pictures into illustrations that evoke the styles of manga (Figure \ref{fig:tiktok}), including creating video reports showcasing their evaluations, (3) discussed ideas about justice and fairness in algorithmic systems, and (4) designed new TikTok filters with Effect House (TikTok’s filter development software). One outcome of the workshop was a method formalizing five steps of algorithm auditing—generating a hypothesis, generating systematic and thoughtful inputs, running the test and tracking input/output pairs, analyzing data, and reporting the findings ~\cite{morales2025learning}, which helped frame later activities, including a Spring 2024 workshop with younger STEM Stars and a Spring 2024 classroom workshop exploring algorithm auditing with youth-created physical computing projects. While all three workshops helped develop curricular ideas for algorithm auditing, we recruited the more senior youth from the EAP workshop to serve as advisors on a lesson design PD session. 

\subsubsection{Algorithm Auditing Lesson Design (AALD) Workshop with Teachers}
Our developing activities began in July 2024, when we recruited six public high school teachers with computer science (CS) teaching experience to participate in the Algorithm Auditing Lesson Design (AALD) workshop. The teachers and researchers gathered for a week-long (30-hour) PD workshop to introduce and explore algorithm auditing and to outline formal lessons, with plans to design more detailed lesson plans during Fall 2024. On the first day, the teachers learned about the five steps of auditing, applied the steps on TikTok filters in a similar manner to the EAP workshop within a shortened time frame of three hours, listened to lectures from HCI researchers on cutting-edge algorithm auditing projects, and reflected on what algorithm auditing might look like in their classrooms. 

The \textit{Teen Advisory Session} took place on the second afternoon of the AALD workshop, after the teachers finished their TikTok algorithm audit reports, with the awareness that they would be presenting these reports for constructive feedback with the teens in the afternoon. In the next sections, we describe explicit preparations with both the teens from the EAP workshop and the teachers in the AALD workshop. 

\subsubsection{Preparing for the Teen Advisory Session: Teens}
\label{sec:prepteens}
Prior to the Teen Advisory Session, Bri and Matías met with the three teens who had chosen to participate to prepare them for meeting the teachers. Following a strong tradition in PD of sharing food ~\cite{roque2023imagining, bonsignore2016traversing}, one meeting took place over ice cream one week prior to the session, and a second meeting included lunch immediately before meeting the teachers. The meals provided a casual and comfortable space to focus on renewing relationships between the teens and researchers, foregrounding the teens’ recollections of their experiences during the EAP workshop months earlier, and eliciting reflections on what they thought about algorithmic bias and justice months after the workshop. 

Matías began the first meeting by introducing the teens to the novel role of being advisors to adults: “You will be experts, talking to teachers." Bri reminded the youth about the content of the EAP workshop in which they had participated. Following this, Matías shared recordings from the EAP workshop as well as research papers produced based on that experience. He asked the youth to watch the recordings and discuss their experiences. Fascinated by the impact their fun work had created, the youth reflected on their everyday practices with TikTok and how they had refined their ideas about filter behaviors through iterative and systematic evaluation. Matías informed the youth that, during the workshop, they would be providing feedback on the teachers’ audit reports. Bri then asked questions such as, “If you were going to give advice to someone evaluating a TikTok filter, what would it be?" to help the youth think of feedback they might share during the advisory session.

During the second meeting, Bri, Matías, and the teens ate Indian food---at the request of the teens---for lunch together. The goal of this meeting was to remind the youth that they had valuable expertise and reassure them that the teachers were eager to learn from them. During lunch, the youth talked about how they had spent their summers, though they also confided to Bri that they were nervous about meeting the teachers. 

\subsubsection{Preparing for the Teen Advisory Session: Teacher-Designers}
After the teachers finished their TikTok filter auditing reports in the form of short slide shows, Clara, who had an extensive relationship with these teachers, guided them in reflecting on potential takeaways of algorithm auditing for lesson designs: considering student relevance, brainstorming potential lesson ideas for their own classrooms, and designing preliminary outlines for lessons. In addition, Clara and Zia prompted the teachers to think of questions they had for the teens for the afternoon Teen Advisory Session. The emerging ideas included whether the teachers’ audit reports would be interesting for teens, what kinds of experiences the teens had in conducting audits (how long it took, what they learned, how they came up with hypotheses), and about other kinds of systems to audit beyond TikTok, which is banned in most schools, such as music recommendation platforms or search results on local attractions. The researchers sought to highlight the experience the teens brought to the upcoming session.

\subsubsection{Teen Advisory Session}
\begin{figure}[h]
\centering
\includegraphics[width=\linewidth]{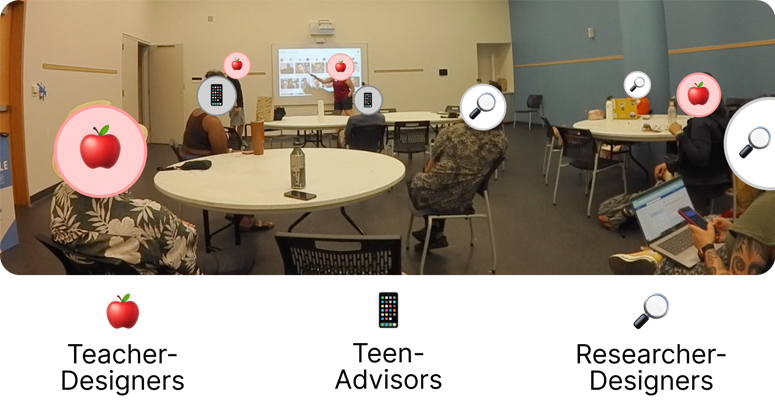}
\Description[A picture of the Teen Advisory Session]{A picture of two (anonymized) teachers presenting their auditing projects, and two other teachers, two teens, and three researchers watching the presentation}
\caption{Two teachers presenting their auditing projects at the Teen Advisory Session.}
\label{fig:session}
\end{figure}

All participants met for two hours at the science center, where the teens did their STEM Star and EAP activities, prioritizing familiarity for the teen-advisors. In contrast to ~\citet{coenraad2022using}, where students were intentionally placed away from adults, the teens were spread out across different tables, seated among the teachers and researchers (Figure ~\ref{fig:session}). The Teen Advisory Session began with a shared viewing of the same video the youth watched in preparation for this workshop: a recording of a previous workshop where youth created TikTok filters using Effect House and conducted peer audits of the filters. Then, the teachers presented their auditing projects in pairs (Table ~\ref{tab:presentation}), with the teens providing constructive feedback and critiques after each.

After the presentations and feedback, participants engaged in an open discussion where the teachers asked questions, shared their emergent curricular ideas they previously brainstormed, and researchers occasionally prompted teens to share some of their experiences and reflections on algorithmic bias, algorithmic justice, and algorithm auditing. Section ~\ref{sec:sharedknowledge} elaborates on themes from this discussion.

\begin{table*}[t]
\caption{Teacher-designers' auditing projects (hypothesis, input images, and results)}
\label{tab:presentation}
\begin{tabular}{@{}llll@{}}
\toprule
\textbf{Presenters} &
  \textbf{Hypothesis} &
  \textbf{Input Images} &
  \textbf{Results} \\ \midrule
\begin{tabular}[t]{@{}l@{}}Phoebe \\ \& Jesse\end{tabular} &
  \begin{tabular}[t]{@{}l@{}}People wearing blue\\ will generate male\\ characters, and wearing\\ pink will generate\\ female characters using\\ the Studio Ghibli filter\end{tabular} &
  \begin{tabular}[t]{@{}l@{}}13 images with blue clothing,\\ 13 images with pink clothing\\ (but only managed to test \\ “a few”)\end{tabular} &
  \begin{tabular}[t]{@{}l@{}}If the clothing is not gender-\\ specific, the filter sometimes \\ changes the gender of male-\\ presenting images to female-\\ presenting images and vice \\ versa. Hairstyle also affected \\ gender results\end{tabular} \\ \midrule
\begin{tabular}[t]{@{}l@{}}Liam\\ \& Leta\end{tabular} &
  \begin{tabular}[t]{@{}l@{}}More masculine-\\ presenting images will\\ result in having glasses\\ than feminine-presenting \\ images using the \\ anime/manga filter\end{tabular} &
  \begin{tabular}[t]{@{}l@{}}3 masculine-presenting inputs\\ without glasses, the same 3 \\ masculine-presenting inputs \\ with glasses, 3 feminine-\\ presenting inputs without \\ glasses, the same 3 feminine-\\ presenting with glasses\end{tabular} &
  \begin{tabular}[t]{@{}l@{}}The filter removed glasses \\ from 33\% of masculine-\\ presenting input images and \\ 100\% of feminine-presenting\\ input images\end{tabular} \\ \midrule
\begin{tabular}[t]{@{}l@{}}Angela\\ \& Tessa\end{tabular} &
  \begin{tabular}[t]{@{}l@{}}Facial expressions on\\ masculine-presenting\\ images will be animated\\ differently than the same\\ expressions on feminine-\\ presenting images using \\ the anime/manga filter\end{tabular} &
  \begin{tabular}[t]{@{}l@{}}5 masculine-presenting inputs \\ with different expressions \\ (open mouth showing teeth, \\ smile showing teeth, smile \\ with closed mouth, tongue out,\\ closed mouth frown) and 5\\ feminine-presenting inputs\\ with the same expressions\end{tabular} &
  \begin{tabular}[t]{@{}l@{}}Feminine-presenting inputs \\ resulted in smaller faces \\ with no teeth and masculine-\\ presenting inputs resulted in \\ facial hair and more instances\\ of teeth\end{tabular} \\ \bottomrule
\end{tabular}
\end{table*}

\subsection{Data Collection}
We collected video data from the teen preparation meetings, the AALD workshop, and the Teen Advisory Session. The teen preparation meeting was recorded on the video conference software Zoom along with a screen recording of the teens watching videos from the EAP workshop. We recorded both the AALD workshop and the Teen Advisory Session with two video cameras with bidirectional audio capture (e.g., microphones facing front and back), placed to record the conversations as completely as possible with an emphasis on capturing the teens and teachers. The cameras were recording continuously during both sessions. 

\subsection{Ethical Considerations/Research Approach and Positionality Statement}
Conducting this type of work ethically requires prioritizing the community's needs and engaging extensively with participants. To ensure this focus, we worked with three teens aged 16--18 from traditionally underrepresented identities in computing. Additionally, our team is committed to maintaining a long-standing relationship with the science center, ensuring all engagements are sustainable and respectful; we describe our prior relationship with the STEM Stars program in Table ~\ref{tab:demographic}. We acknowledge the potential risks associated with using tools and technologies developed by TikTok, especially for minors. To address these concerns, participants were provided with project phones, computers, and private TikTok accounts for the activities.

We also collaborated with teachers from districts in two areas in the United States, largely working in public schools with majority low-income students. Our relationships with the teachers stretch 5--10 years over multiple collaborations, and we seek to maintain long-standing relationships with the teachers. During the AALD workshop, we foregrounded teachers’ ideas and curricular interests in design. 

We also recognize that our own identities and backgrounds shape our research approach. Our team represents at least five racial/ethnic identities, three gender identities, and academic expertise in the learning sciences and human-computer interaction (HCI). Most of our team resides in the same location as the teens and the science center, while Clara lives in the same location as three of our teachers and has been in their classrooms repeatedly over a decade. Our qualifications---including experience conducting expert audits, teaching high school students, and designing learning environments---enabled us to carry out this study responsibly and effectively. Of note, the first author of this paper was not present during this part of the workshop and indeed was new to the study. This provided some fresh perspectives on the back stage of design, the interactions between teen advisors, researchers, and teachers in analysis.

\subsection{Analysis}
Prior to the analysis, the recorded videos of the student preparation session, teacher preparation session, and the Teen Advisory Session of this study were descriptively transcribed. The analysis was done by the first author, with the second author’s feedback and the larger team’s consensus, iteratively through four rounds of thematic analysis ~\cite{saldana2016coding, braun2006thematic}. In the first round, the coder reviewed the transcriptions to gain familiarity with the data, highlighting whether each turn-of-talk was an inquiry or insight and finding general content-level themes within conversations. These initial themes included technical knowledge, cultural knowledge, auditing as a learning experience, youth interests, and personal experiences with AI systems. The general themes formed the basis for the second round of coding and to generate a codebook to systematically document the types of knowledge the youth contributed. We modified the codebook to explicitly focus on youth contributions. The resulting themes included \textbf{technical knowledge}, both in the use of technology (e.g., TikTok, phone cameras) and algorithm auditing practices, \textbf{personal relevance} in algorithmic systems and audits, in which the youth shared both their \textit{personal} interests in algorithmic systems and reasons why \textit{other} teens might become interested in algorithm auditing,  and \textbf{potential outcomes of auditing experiences}, particularly when the youth discussed how they changed their thinking about their everyday experiences with algorithmic systems and how they have applied what they learned since the EAP Workshop. 

The first two authors discussed these themes and, following feedback from the team, iterated on a third round of coding, focusing on how the teachers and researchers dialogically positioned (i.e., in their conversational moves) ~\cite{holland1998identity} the youth in their inquiries to understand how the teachers and researchers encouraged the teens to take on the role of advisors. The teachers' and researchers' turns-of-talk were organized into the following themes: \textbf{explicit recognition of youth expertise} in technology and youth interests, \textbf{encouragement of conversation} through direct invitations for the youth to speak and elicitation of prior experiences, and \textbf{finding common ground with youth} based on shared experiences and peer-level conversations. We organized and added the preliminary themes to the codebook, and the coder, with the consensus of the team, proceeded with a final round of coding. The results from the second round of coding address the first research question, and the results from the final round, coupled with a thorough retrospective analysis of the design of the preparatory meetings and the Teen Advisory Session, address the second research question. Sections of the descriptive transcriptions were also transcribed verbatim to highlight significant interactions that emerged in the session.

\section{Findings}
In this section, first, we describe the types of knowledge and practices shared by the teens when situated as advisors. Then, with an eye toward the “back stage of design” ~\cite{bodker2017tying}, we describe ways in which the preparation, structure, and emergent facilitation strategies encouraged the teens to take on the role of advisors. 

\subsection{Shared Knowledge and Practices from Teen Advisors}
\label{sec:sharedknowledge}
When situated as advisors, the teens shared knowledge and practices by demonstrating expertise in auditing practices and in using TikTok, identifying personal relevance for the design of learning activities, and conveying what they learned from the algorithm auditing activities.

\subsubsection{Teens’ Constructive Feedback on Audit Designs}
\label{sec:feedback}
One area where the teens made distinct contributions as advisors included insights into the technicalities of conducting algorithm audits of TikTok filters, as well as knowledge from prior experiences in developing and evaluating audits.

As prior frequent users of the TikTok social media platform and as part of the EAP workshop, the youth had expertise to share on the functionalities and limitations of using TikTok filters. Many of the comments they made related to this expertise, for instance, avoiding issues that might affect filter outputs---glare or lighting, appropriate non-distorting backgrounds, limiting the number of people in a filter (since filters generally handle only one or two people)---and providing logistics knowledge about how to apply filters to pictures in the camera roll. As an example, when providing feedback on the teachers’ audit presentations, Eric correctly deduced that the teachers took pictures of another device’s screen, resulting in reflective glare: “If it’s a picture on another screen, the glare can definitely impact the results and rendering.” This kind of feedback drew on both the teens’ robust prior usage of TikTok and cell phone cameras ~\cite{vogelstein2025rapid} as well as their experiences conducting audits in the EAP workshop, focusing on creating high-quality data inputs that could be used in an audit. 

In addition, the teen advisors provided constructive feedback on the suitability of the datasets for the teachers’ specific hypotheses. For instance, Leta and Liam (both teachers) examined whether gender influenced how often an anime filter retained glasses in the image. In providing constructive feedback, Nadine complimented the internal consistency and clarity of the input data (i.e., three feminine-presenting and three masculine-presenting people with glasses and the same people without), noting, “I loved the consistency. I thought when you used pictures of a person with glasses and then [the same person] without glasses […] that was great that you kept the control.” In this, Nadine pointed out how the two teachers avoided unnecessary data variation by using the same people with and without glasses. Eric added that a larger pool of input data would have been helpful to answering the questions while encouragingly noting that “the hypothesis was really unique.” Here, the two teens provided clear, constructive feedback: encouraging Leta and Liam with positive feedback on what they did well and proposing ways to improve their audit. At other times, the teens highlighted how fluctuations in lighting, articles of clothing, and the quantity of people in the input data were unfavorable for maintaining data consistency when testing the teachers’ specific hypotheses. In these comments, the teens demonstrated their ability to provide clear, constructive feedback to the more novice teachers. 

\subsubsection{Unique Insights on Teen Perspectives on Auditing Algorithmic Systems}
\label{sec:insights}
Given that the goal of the overall design was to create a set of algorithm auditing lessons to implement in U.S. high school CS classrooms, the dialogue between teachers and teens became critical in informing curriculum design. These exchanges were particularly valuable in soliciting the teens’ perspectives on the teachers’ proposed ideas for algorithm auditing lesson topics. Through structured conversations, the teens not only validated initial ideas posed by the teachers but also expanded them, highlighting the importance of personal relevance and youth interests. 

During the advisory session, the teachers sought teens’ feedback on whether some in-progress curricular ideas felt interesting, relevant, or engaging. One idea involved auditing music recommendation systems like Spotify, to which all three teens expressed personal interest, with Mel expanding the idea by suggesting that they could audit Spotify’s Radio playlist generation algorithm. A broader discussion ensued after this; one teacher (Leta) raised the question of whether auditing shopping recommendation systems like Amazon’s might be more engaging than music recommendation systems. Mel and Eric shared differing opinions, with Mel interested in shopping platforms and Eric more intrigued by music. Leta concluded, “So that could be different choices! We need choices!” The conversation then shifted into more design ideas for the lessons. In another interaction, Tessa, who taught in the same Mid-Atlantic city as the teens, presented a lesson idea to experiment with Google searches of city attractions local to each classroom’s location, noting that her own students are passionate about their city. The teens supported this idea, with Mel specifically adding that this could further focus on local attractions for teens (i.e., young people), noting that there’s not a ton for teenagers to do in the city and how it would be interesting to see where they could go “beyond the tourist focus.” These interactions demonstrate some of the emergent productive dialogue, providing insights to teachers on whether teens might be interested in certain ideas, expanding those ideas, and generating new ideas. As advisors, the teens provided insights unavailable to the teachers simply based on age and generational knowledge. 

The teen advisors also shared insights into why they found working with TikTok meaningful. For example, Nadine described her initial hesitation about using TikTok in a STEM learning context. She noted that, prior to the EAP workshop, she spent up to nine hours a day on TikTok but had never considered it in a more critical fashion. What engaged her most was realizing, 
\begin{quote}
“Oh wait, this is at my fingertips? The same TikTok that’s on my phone is being used for research. When you use TikTok, you don’t really think about it. But when you sit down and really analyze it, you realize how important it is.” 
\end{quote}
While the teachers were already aware of TikTok’s prominence in youth’s lives, Nadine provided this nuanced perspective of how the contrast between TikTok for leisure and for research could motivate deep engagement for youth. Mel added another layer to the discussion, sharing how evaluating the behavior of a popular TikTok filter alongside peers her age drew her into the auditing process. She explained that she had always dismissed filters that didn’t work for her as “janky, because it’s a new filter,” but through the workshop, she began thinking critically about “why that may be.” This provided novel insights to the teachers who, by their own acknowledgment, were largely unfamiliar with using TikTok filters prior to that week. The teens’ reflections highlighted the importance of leveraging tools familiar to youth as a way of empowering them to critically analyze systems they regularly use. 

\subsubsection{Teen Reflections on Prior Auditing Experiences}
\label{sec:reflectionsonaudit}
A key component of the conversation between the teen-advisors, teachers, and researchers circled around what the teens had learned, valued, and found interesting about their experiences with algorithm audits and designing their own TikTok filters. At the end of the session, the teens articulated how the EAP workshop shifted their perspectives on algorithmic systems. They described developing a heightened awareness of how algorithmic systems shape their lives, noting how their new understanding led them to question these systems they previously took for granted. For instance, Mel talked about her experience with a popular filter that turned people into manga characters. She noticed that “there is little to no representation of darker-skinned people, let alone Black women,” on the filter, referring to how the filter often changed the skin tone of people with darker skin. However, through her experimentation with Effect House, she was able to create filters that worked for Black women, concluding, “It’s possible; it can be done. It’s just, like, nobody is trying to do it.” Nadine continued Mel’s concerns, stating that these big companies and “geniuses” know what they are doing, yet they are still excluding people of color. She later noted, “There is error in the thing that should have no error.” The teens’ observations reflected their changed perspectives based on their experiences auditing and creating AI/ML filters.

The teens also shared concrete ways that they had changed everyday interactions with algorithmic systems since the EAP workshop. For example, they described how they actively shaped recommendation systems on platforms like YouTube. Mel mentioned her new awareness of browser cookies, actively observing how advertisements often reflected recent website visits. Eric added, “It’s made me a lot more aware of different kinds of content I’m consuming and different things I’m clicking on.” He recounted deliberately shaping “the perfect algorithm” on YouTube by clicking on every video he wanted to watch, emphasizing how quickly the system personalized its recommendations. Lastly, Nadine described newfound caution in identifying AI-created materials online, adopting a more critical view than before when thinking about whether “ads and stuff” were created or promoted by AI systems, "because it could really just be someone from any different country. You never know.” 

The teen-advisors’ reflections gave teachers and researchers insights into how learning activities with algorithm auditing or creating TikTok filters could help high school youth develop more critical AI literacies. Angela’s response to the teens’ reflections captured a key insight for the designers:
\begin{quote}
“I think these kinds of activities are super beneficial […] you instantly gain awareness of what is happening even if you don’t know the technical stuff. From becoming a passive [user of] technology, you’re thinking about it a little more […] question it a little bit.” 
\end{quote}
The sequencing of these conversations---occurring months after the EAP workshop---provided teachers and researchers with insights into the potential of learning about algorithm auditing. Listening to the teens reflect as advisors on the scoping activities gave insights into what the teens learned and how the learning could impact thinking beyond the classroom.

\subsection{Back Stage Design for Situating Youth as Advisors}
As ~\citet{iversen2012scandinavian} point out, power relationships color interactions during PD, especially the power differential between adults (in this case, teachers and researchers) and teens\footnote{We note that there is also a major power differential between researchers and teachers, but since the emphasis of this paper is on the teens in advisory roles, suffice it to say here that the researchers sought to prioritize teachers’ voices, ideas, and leadership throughout the PD session.}. In this section, we share analysis of the in-the-moment conversational moves from teachers and researchers that helped to situate youth as advisors during the session. We also highlight our retrospective analysis of the “back stage design” that helped prepare teen-advisors and teachers for this collaborative meeting, including pre-session meetings, choices of session activities, and session sequencing.

\subsubsection{Explicitly Positioning Youth as Experts}
The decision to use TikTok as a platform for learning about algorithm auditing in both the EAP and AALD workshops aimed to make the learning experience more relevant to youth ~\cite{morales2025learning, vogelstein2025rapid} and created an opportunity for the teens to act as advisors. The teens shared valuable advice and insights, drawing from their experiences as TikTok users and as teenagers. In addition, the teens had more experience conducting audits on TikTok filters---spending eight hours across two sessions---compared to teachers, who only conducted three hours of auditing before the advisory session. With these choices, the teachers and researchers relied on and learned from the teens’ knowledge in meaningful ways.

To challenge the traditional teacher-student dynamic from the outset, we foregrounded interactions with TikTok filters and auditing at the beginning of the Teen Advisory Session, emphasizing the teens’ expertise. Beginning the session with the recording from the EAP workshop highlighted the teens’ experience with Effect House, a platform entirely unfamiliar to the teachers. Following the video, Liam acknowledged the teens’ familiarity with the technology by directly asking the teens, “Could you explain how to make filters on TikTok?” Eric responded by outlining the process of creating a filter on Effect House and further describing some complications he faced with the platform. This opening interaction immediately positioned the teens as knowledgeable contributors, reinforcing their role as advisors and setting the tone for the session.

After this introduction, the teens provided feedback on the teachers’ auditing projects, intentionally positioning the teens as knowledgeable experts and evaluators. During their auditing presentation, one teacher (Jesse) expressed discomfort with using TikTok, stating, “[w]e had huge technical difficulties rendering in TikTok,” and “this is actually the first time we’ve ever used it!" framing the teachers as inexperienced and open to more expert feedback. It was in response to these comments that Eric, Mel, and Nadine provided constructive criticism as discussed above (Section ~\ref{sec:feedback}). Thus, positioning the teens to provide feedback on the teachers’ presentations, along with the teachers’ drawing attention to their own inexpertise, supported the teens in advisory roles. 

Beginning the Teen Advisory Session with these activities helped clearly position the teens as more expert TikTok users and auditors, supporting the teens’ roles as advisors with teachers as learners. Conversational moves by both the teens and teachers helped reinforce this complementary positioning in amicable ways. Next, we step back to consider how the preparatory meetings helped frame these roles.

\subsubsection{Eliciting Prior Experiences and Encouragement of Conversation}
As described in Methods (Section ~\ref{sec:prepteens}), we intentionally held a pre-advisory session meeting to prepare the teens to take on the role of advisors. Here, we consider how the preparations with teens and in-the-moment facilitation in the advisory session helped connect across vertical levels of PD, especially between the teens and teachers. 

In the youth's preparatory meeting, the teens displayed initial confusion on the type of feedback they would provide, with Eric questioning the types of critiques they should be giving. Acknowledging their hesitance, Bri suggested showing the teens examples of the types of presentations they would be critiquing during the Teen Advisory Session. Together, Matías, Bri, and the teens revisited previous auditing presentations from their peers in the EAP workshop. The teens categorized feedback they could bring up with the teachers, including comments about lighting in the input data, variability of the number of people in the input data, and limitations of the filters on darker skin color. Indeed, they raised all of these points during the later advisory session. Thus, the preparatory meeting not only clarified the teens’ role as advisors but, at a distance of eight months since the EAP workshop, reminded them of knowledge built during the workshop and elicited new meta-reflections on insights that might help people new to auditing TikTok filters. The direct link between this discussion and the teens’ feedback on teachers’ auditing presentations demonstrates the importance of this preparatory meeting. 

In addition to the pre-session meetings, the researchers used various in-the-moment facilitation strategies that further encouraged the teens’ participation. One strategy was to issue general probing questions to the teens that directly invited their insights. For instance, after the first presentation of teacher-created audit reports, Matías broadly asked if anyone had any thoughts. The two presenting teachers (Phoebe and Jesse) added to this question by asking the participants how they could make TikTok filters have more consistent outputs. After a short lull, one teacher suggested that they could edit the pictures to control for external variables. To engage the youth in this conversation, Matías directly asked the teens, “[STEM Stars], what do you think about this?” This direct invitation, alongside positioning the teens with their title of STEM Stars, elicited Mel’s critique on the variation in the number of people in the teachers’ input images and Eric’s comment on the glare from taking pictures of another device’s screen (Section ~\ref{sec:feedback}). From this point forward, the teens eased into their role as advisors, consistently voiced their feedback, and shared expertise after each of the teachers’ presentations. Other instances of emergent conversational moves also highlighted the teens’ knowledge, such as when Clara posed questions like, “What helped you get into that ‘this is kind of important’ zone?” and “How has auditing changed your perception of the world?” which prompted reflective insights and further enriched discussions.

A second facilitation strategy involved a researcher calling out specific past experiences with the teens that were relevant in the moment. For example, building on his established relationship with the teens across multiple workshop sessions over three years (Table ~\ref{tab:demographic}), Matías posed targeted questions that evoked specific memories and insights that might otherwise have remained unmentioned. For instance, Matías recalled Mel’s and Nadine’s “strong feelings about evaluating TikTok” and invited the teens to talk about how it felt to “see the other side” of creating the filters. Consequently, Mel shared her critical analysis of the Manga filter that could not accurately generate Black women, and Nadine shared her concern about big companies that were excluding people of color. Similarly, at the end of the Teen Advisory Session, Bri described noticing the STEM Stars “confidently talking about what they were doing” in the EAP workshop because they were already so familiar with TikTok. Following this, Matías prompted Nadine to share about her evolving perspective on TikTok, saying, “[Nadine], you had a strong reaction from going to “TikTok is a waste of time” to taking TikTok seriously. Nadine elaborated that thinking deeply about TikTok as a platform that could be critically analyzed helped her engage with the workshop (Section ~\ref{sec:insights}). Throughout the advisory session, these elicitations of the teens’ experiences created opportunities for the teens’ substantive contributions. These interactions highlight how the longitudinal relationship between a researcher and the teens served as a bridge, connecting threads across the session to prompt the teens’ nuanced reflections.

\subsubsection{Engaging in Authentic Dialogue and Collective Reflection}
Authentic dialogue is more than exchanging ideas---it is a process of mutual engagement where participants listen, respond, and build on each other's perspectives in a way that fosters shared understanding. In Section ~\ref{sec:feedback}, we discussed that the teens had more experience with algorithm auditing than the teachers, which helped solicit feedback and constructive criticism from the teens. The teachers and teens, nonetheless, shared common experiences conducting audits on TikTok filters. These shared experiences were instrumental in emphasizing the validity of the teens’ perspectives and easing existing power dynamics between teens and teachers. Consider when Mel provided feedback on Tessa and Angela’s auditing presentation (gender differences in filter outputs), pointing out how the filter they audited changed the skin tone of the people in the input images and even “gave up” on capturing her braids. Tessa related with Mel, stating, “I like the term ‘give up.’ It felt similar for me.” With this simple statement, the teacher validated Mel’s frustration and communicated a shared feeling about the filter’s limitations. 

Further, as conversation shifted from feedback on presentations to general reflections on algorithmic bias, dialogue bounced across all participants, who found common experiences from auditing and everyday observations with algorithmic systems in their daily lives. For example, toward the end of the session, Clara asked how auditing had changed participants’ views of technology. Eric noted he now saw technology as “always learning” from what it observes. Matías argued that “just because there are products that work, does not mean that they work for everyone in the same way,” further exclaiming, “But I also think that’s interesting, because if there are flaws, there’s always space for improvement!” This critical engagement with Eric’s perspective, while affirming and adding to his statement, sparked an extended discussion among many of the participants---the three teens, three teachers (Tessa, Leta, Jesse), and two researchers (Matías and Clara). Conversation moved from one person to the next as they explored topics like feeling siloed on the internet (Mel, Tessa, Leta, Eric), feeling more cautious of AI-generated content (Nadine), potential regulations to limit recommendation systems (Clara), and the harmful impacts of social engineering (Jesse). 

To illustrate, when Mel shared her concerns about YouTube tracking her browser cookies to show related advertisements, Tessa seconded her, elaborating, "Yeah, exactly. Personally, I get concerned that I'm getting siloed." Leta furthered the discussion, exclaiming that, "It makes me angry. I really live in a bubble. There's so much more I'm not getting exposed to." All of this resulted in Eric sharing his experience curating his YouTube algorithm and his unease with the platform's addictive qualities (Section ~\ref{sec:reflectionsonaudit}). The progression from unidirectional exchanges during presentation feedback to collaborative discussions highlights how the session began to foster an environment where teens’ insights were not only acknowledged but actively built upon. By considering the teens as knowledgeable contributors and engaging with their ideas as substantive and worth debating, the participants demonstrated how authentic dialogue can bridge vertical and experiential differences, creating a shared space for reflection and collective learning.

\section{Discussion}
This paper contributes a new role in CCI: teens as advisors in PD. Through a case study, we examined how teens worked alongside teachers and researchers in an advisory capacity during the design process. In the discussion, we review the role of youth as advisors, what we learned from developing these participatory design sessions, and how this role can contribute to the design of learning experiences with emerging technologies. 

\subsection{The Role of Teens as Advisors: Legitimate Youth Contributions in Participatory Design}
Our analysis reveals that it is indeed possible and even desirable to situate teens as advisors in the PD process. In this project, teens brought particular expertise with emerging technologies, due to their prior and extensive experiences with algorithmic systems both in everyday life and in an earlier exploratory workshop. During the two-hour advisory session, the three teen-advisors actively participated in productive dialogue, contributing to new directions for lesson designs on algorithm auditing. The teens and teachers exchanged perspectives and ideas, listening and responding thoughtfully to each other in a reciprocal, dynamic conversation. In these conversations, the teens leveraged their extensive technical expertise to provide detailed feedback about teachers’ data and algorithm auditing results. Further, they shared recommendations for other familiar algorithmic systems, such as Spotify Radio, and reflective insights into what they learned from the algorithm auditing workshops, providing valuable information for teachers and researchers for the design of classroom learning activities. Engaging the teens as advisors during a strategically placed two-hour session answered some past challenges on involving teens in PD, a group that has been relatively overlooked in CCI research ~\cite{fails2013methods, iversen2013understanding, smith2023research}. The session provided a time-limited way to involve teens in PD, an alternative to accelerated or focused design sessions ~\cite{pazmino2015designing}. 

Assigning the role of advisor also challenged traditional power differentials between young people and adults, elevating youth contributions in the design process beyond the role of informant, while still allowing for limited time investment by the teens. Situating youth as advisors connects with the protagonist reframing proposed by ~\citet{iversen2017child}, which “serves the purpose of empowering children to take an active role in developing technology in making decisions about technology in their lives” (p. 34). Assigning the role of advisor helped to “create legitimacy and ownership” ~\cite{iversen2017child}. It is here where the distinction between the roles of informant and advisor becomes clear. While in either role, participants provide insights and feedback, as advisors, their expert knowledge is being explicitly acknowledged---which led one teen to state during the workshop session that for the first time their experience with TikTok was being validated.

Situating teens as advisors in this study also provided an avenue to involve them in different roles across distinct stages of the broader PD project ~\cite{landoni2018comparative}. In long-term PD initiatives, such as the scalable design of classroom curricula, participants often shift in and out of the process. Their availability and capacity may also vary throughout different dimensions (scoping, developing, scaling) of the project’s development. In this case, the teens initially participated as design partners during the earlier scoping stage of the project ~\cite{morales2025learning, vogelstein2025rapid}, engaging in algorithm auditing with TikTok filters. Eight months later, they leveraged their prior experiences and reflected on their learning to serve as advisors during the early developing stage. Their ability to draw on their past experiences---evidenced by their feedback on algorithm auditing and their advice to teachers and researchers---demonstrates the potential of building long-standing relationships with youth to support their changing roles in PD.

This points to an opportunity to further explore how youth participants in earlier stages of a project, whether as users, testers, informants, or design partners, can transition into advisory roles in later phases. By fostering long-term relationships and enabling pathways for sustained engagement, PD can better integrate youth perspectives at multiple points throughout the design process. Just as these teens transitioned from the role of design partners---developing the five steps of algorithm auditing ~\cite{morales2025learning}---to advisors in the Teen Advisory Session, future studies could also provide examples of ways that teens who participate in roles as users, testers, informants, or design partners in earlier stages of a PD project can build on those experiences to serve as advisors in a later phase of the project. Furthermore, while our research focused on teens as participants, it might be worthwhile to explore whether younger participants can be placed in an advisory role and contribute to the design.

\subsection{Identifying Back Stage Design Efforts: Long and Short Term}
Following Bødker, Iversen, and Dindler’s ~\cite{bodker2017tying} call to illuminate the messy “back stage of design,” we identify both long- and short-term efforts that helped set the stage for teens to serve as advisors on this project. First, we involved teens in different stages of the design, initially as design partners in exploring scoping activities with algorithm auditing and later as advisors to teachers and researchers working to create classroom lessons. This further built on even longer-term relationships with the science center staff and the STEM Stars program, and earlier work on other projects with these same teens and their peers. All this helped create trust between the youth, staff, and researchers, which was essential in asking the youth to step into a new role---which they initially expressed nervousness about---of advising teachers.

We also built on long relationships with the teachers, which were critical for selecting teachers who we knew were likely to support youth agency and ideas. These relationships also created a foundation of trust with teachers when they were invited to step into the role of being advised by teens. This was possible due existing infrastructures of collaboration with the STEM Stars and the teachers that could be leveraged to bring different communities together \cite{disalvo2012communities, dantec2013infrastructuring, poderi2018participatory}. All of this reaffirms literature that argues that PD requires complex and long-term engagement and working across vertical levels of power ~\cite{smith2018participatory}. Indeed, asking participants to work across typically stratified roles (student---teacher---researcher) and different types of institutions (science center---public schools---higher education) required trust built over years of relationships. We agree with \citet{disalvo2012communities} on the importance of long-term engagement with communities, but argue that PD should also facilitate opportunities to bring different communities of relevant parties together in ways that recognize the expertise and value of those parties (i.e., in our case high school youth) that are often overlooked in the design process.

Our analysis also revealed the importance of horizontal back stage work: the micro-dynamics necessary for each relevant party to participate ~\cite{bodker2017tying}. Here, the preparatory meetings conducted with youth supported them in reviewing their prior audits during the EAP workshops and helping them think through what they \textit{already} knew. Doing this over meals with the teens’ choices of food and in the familiar setting of the science center helped prioritize the teens’ sense of comfort. The preparatory meeting with the teachers helped foreground questions and topics for learning, encouraging the teachers’ positioning as learners before the advisory session. In addition to the meetings, the choice of TikTok filter audits as the first activity of the session provided a starting point for conversation where the teens had much more experience than the teachers. 

At the same time, in-the-moment facilitation moves by teachers and researchers encouraged the teens to speak up about their experiences. Again, the long relationships with the teens meant that the researchers had knowledge about the teens’ experiences, which they invited the youth to share. This opened up conversations where the teens offered critiques, added specificity to the teachers’ ideas, and generated new considerations about algorithm auditing for classroom implementation. These micro-dynamic interactions helped empower youth---in-the-moment---with agency, ensuring their voices and opinions had a place of priority in the discussion ~\cite{bodker2017tying, iversen2012scandinavian}. 

\section{Limitations}
In this paper, we focused on the analysis of the two-hour, transcribed video recorded interactions between teen-advisors, teacher-designers, and researcher-designers to understand what knowledge and practices they would provide on in-process lesson plan designs for algorithm auditing in the classroom. With this observational data, we focused on the roles youth and teachers took in practice. Being an advisor is seldom just a one-way street where designers benefit from the wisdom of the feedback; advisors also benefit and learn from the process. Future studies should include interviews capturing the different parties’ reflections about taking on these unusual roles or what they may have learned during the process. Such analysis could provide useful insights into youths’ concerns and motivations about advising. Further, this study of a single session focused on the specific area of algorithm auditing. We look forward to additional PD studies that consider teens as advisors in other domains, perhaps at multiple stages of design, with insights into further facilitation strategies and preparation that might support these exchanges. 

\section{Conclusion}
We propose a new role for youth in participatory design: advisors to researchers and teachers. We observed that youth offered constructive and technical feedback on algorithm auditing lesson plans, drawing on both their substantial expertise gained in everyday life as well as insights from previous workshops in earlier scoping stages of design. We also examined the back stage elements of the PD process that addressed the power differential between participants and helped youth step into the roles of advisors. Here, we gained insights into how specific choices---including the technology and topic selection (e.g., algorithm auditing on a familiar social media platform), location decisions, cultivation of long-term relationships, and in-the-moment facilitation strategies---all helped give voice and agency to youth.

\section{Selection and Participation of Children}
We recruited teens already enrolled in a STEM program at a science center in a city located in the Mid-Atlantic United States. The teens were invited by the organizer of the STEM program to participate via email. Parents received consent forms prior to the study, which included a brief explanation of the research, and the teens assented to their participation. Research protocols and data collection methods were approved by the IRB board of the University. Teens received hourly compensation for their time as well as meals during sessions.

\begin{acks}
The analysis and writing of this paper were supported by National Science Foundation grants \#2333469 and \#2342438. Any opinions, findings, and conclusions or recommendations expressed in this paper are those of the authors and do not necessarily reflect the views of NSF, the University of Pennsylvania, or The Franklin Institute. 
\end{acks}

\bibliographystyle{ACM-Reference-Format}
\bibliography{references}

\end{document}